\documentstyle[12pt]{article}
\begin{document}
\title{\large \hspace{10cm} ITEP-11/00 \\ \hspace{10cm} March 2000 \\
\vspace{1cm}
\LARGE \bf
On decay of large amplitude bubble of disoriented chiral condensate}
\author {T. I. BELOVA
\\
{\it Institute of Theoretical and Experimental Physics,}\\
{\it B.Cheremushkinskaya, 25, Moscow, 117259, Russia}\\
\\
V. A. GANI\thanks{E-mail: gani@heron.itep.ru}{\,}
\\
{\it Moscow State Engineering Physics Institute (Technical University),}\\
{\it Kashirskoe shosse, 31, Moscow, 115409, Russia}\\
{\it and}\\
{\it Institute of Theoretical and Experimental Physics, Russia}\\
\\
A. E. KUDRYAVTSEV \thanks{E-mail: kudryavt@heron.itep.ru}
\\
{\it Institute of Theoretical and Experimental Physics, Russia}\\
}
\date{}
\maketitle
\vspace{1mm}
\centerline{\bf {Abstract}}
\vspace{3mm}

The time evolution of initially formed large amplitude bubble of disoriented
chiral condensate (DCC) is studied. It is found that the evolution of this
object may have a relatively long pre-decay stage. Simple explanation of
such delay of the DCC bubble decay is given. This delay is related to
the existence of the approximate solutions of multi-soliton type of the
corresponding radial sine-Gordon equation in (3+1) dimensions at large
bubble radius.

\newpage

     In our previous paper~\cite{GKB} we discussed the time evolution
of initially formed DCC bubbles in the simplified chiral two-component
classical sigma-model ($\sigma^2+\pi^2=f_{\pi}^2$). In such model it is
convenient to introduce a new field variable $\phi$: $\pi=f_{\pi}\sin{\phi}$,
$\sigma=f_{\pi}\cos{\phi}$. The equation of motion in terms of $\phi$,
studied in Ref.~\cite{GKB}, has the form
$$
\phi_{tt}-\phi_{rr}-\frac{2}{r}\phi_r+\frac{m^2}{n}\sin{(n\phi)} = 0 \ ,
\eqno(1)
$$
where $\phi\in[0,2\pi]$, $m$ is the mass of the field $\phi$ and $n$ is
integer. If $m=0$, we get the case of unbroken chiral symmetry. If $m\ne 0$,
the chiral symmetry is broken. In the case $n=1$ the theory has the only
vacuum state at $\phi=0$. In terms of the field variable $\phi$ the bubble
of DCC corresponds to the following field configuration: everywhere inside
the spherical symmetrical domain $\phi=const\ne 0$; everywhere outside this
domain $\phi\equiv 0$ (true vacuum, i.e. $<\sigma>=1$, $<\pi>=0$).
The decay of DCC bubble in the model with $n=1$ leads finally to the
formation of a breather-like solution~\cite{GKB,Her}, located at the
center of the initial bubble. Formation of a long-living breather solution
is a typical phenomenon for a wide class of nonlinear systems including
one described by Eq.~(1). It is worth to mention that, because of
nonlinearity of the problem, the mean life-time of large amplitude DCC
bubbles significantly exceeds one of linearized DCC systems with external
sources, see, e.g. Ref.~\cite{Bjorken}.

     The case $n\ge2$ corresponds to the theory with $n$ degenerate
vacua at $\phi=0$, $2\pi/n$, $4\pi/n$, ..., $2(n-1)\pi/n$.
As it was demonstrated in Ref.~\cite{GKB}, the evolution picture of
DCC bubbles in this case depends crucial on the initial amplitude.
The case of small amplitude also leads to the formation of a breather-like
solution. But initially formed bubble of large amplitude behaves in a
different way which is characterized by rather long pre-decay stage with
relatively low radiation.
This first stage of evolution consists in splitting of the shell of the
initial bubble into several concentric shells of smaller amplitudes.
The next step is the interaction of these shells. To the end we observed
emission of the main part of the initial energy of the bubble in the form
of waves of small amplitude followed by formation of a long-living
localized breather-like solution in the center.
This rather complicated picture may be called the delayed decay of
DCC bubble. In this paper we continue studying the process of DCC bubble
decay for the special initial conditions
$$
\phi(r,0)=\frac{2\pi}{1+(r/r_0)^K} \ , \ \ \phi_t(r,0)\equiv 0 \ ,
\eqno(2)
$$
where $K$ is a large and positive number, in our calculations we assign
$K=20$. As it was discussed in Ref.~\cite{GKB}, the evolution picture
of the initial bubble depends crucial on dimensionless parameter $\xi=mr_0$.
In the case of small $\xi<\xi_{cr}\sim 2$ in the model with $n=2$
we observed a prompt decay of DCC bubbles followed by formation of a
breather-like solution. But for $\xi>\xi_{cr}$ we observed splitting
of $2\pi$-shell of the initial bubble into a pair of concentric
$1\pi$-shells. The transition from prompt decay to delayed decay is
clearly seen in Fig.~1. In this Figure we give the energy flux through
the sphere of radius $R>r_0$ in units of the total energy at two typical
moments of time as a function of the parameter $\xi$.
As it is seen from Fig.~1, at $\xi<\xi_{cr}$ the main part of the released
energy is emitted from the region of the bubble during a relatively short
period $T<50$ (in dimensionless units).
But at $\xi>\xi_{cr}$ we observed that only some part of the released
energy is emitted from the region of the bubble during the same period of
time. This phenomenon we call the delayed decay of DCC bubble.
In Fig.~2 we give pictures of the field configurations ($n=2$, $\xi=10$)
at some typical moments of time $t_i$. As it is seen from Fig.~2, further
evolution of the split $1\pi$-shells leads to their secondary
interaction. This interaction is of repulsive character and it takes place
if their radii coincide. After several collisions of $1\pi$-shells the total
configuration transforms into a localized breather-like solution at the
center of the initial bubble.

     The qualitative explanation of the observed splitting of $2\pi$-boundary
of the initial bubble (2) into a pair of concentric $1\pi$-shells may be
the following. Assume, that for sufficiently large radius $r_0$ of the bubble
the term $\frac{2}{r}\frac{\partial\phi}{\partial r}$ in the left-hand side
of Eq.~(1) is getting inessential and may be dropped. Making so, we get from
Eq.~(1) the one-dimensional sine-Gordon equation on semi-axis $0\le r<\infty$.
Solutions of this equation at large $r$ look like solutions of the integrable
sine-Gordon equation. Multi-soliton solutions of the one-dimensional
sine-Gordon equation may be obtained analytically. In particular,
the double-soliton solution for $n=2$ looks like \cite{Rad}
$$
\phi(x,t)=
2\arctan\left(\frac{v\sinh{(mx/\sqrt{1-v^2})}}{\cosh{(mt/\sqrt{1-v^2})}}
\right) \ ,
\eqno(3)
$$
where $v$ is the solitons' velocity at the infinity. At an arbitrary moment
of time $t$ this solution looks as a superposition of two solitons with the
integral of motion
$$
\phi(+\infty,t)-\phi(-\infty,t)\equiv2\pi \ ,
$$
called "topological charge". At $t=0$ solution (3) is a $2\pi$-jump of
characteristic size $x_{char}\sim\sqrt{1-v^2}/mv$. In the relativistic limit
$(1-v)\ll 1$ the size $x_{char}$ is small, $x_{char}\ll m^{-1}$. Because of
similarity of shapes of both the initial condition (2) at $K\gg 1$ and
the solution (3) at $t=0$ in the relativistic case, the time evolution of
both solutions should also be similar at least at small positive $t$.
Looking at solution (3) at $t>0$, we get that $2\pi$-kink splits into a pair
of $1\pi$-kinks moving in opposite directions. Just the same happens
at $t>0$ with the solution of Eq.~(1) with initial conditions (2).
So, we conclude that 3-dimensional $2\pi$-shell behaves like the double-kink
solution (3) at least for small positive $t$.

     Further evolution of both separated $1\pi$-shells looks differently.
The internal $1\pi$-bubble behaves as the usual large amplitude bubble
in the model with $n=1$, studied long ago by Bogolubsky and Makhankov
\cite{Bog}, see also Review \cite{Bel1}. It shrinks and expands again,
emitting part of its energy. The external $1\pi$-shell expands, then stops
and goes back. The maximal radius of this $1\pi$-shell may be estimated
from the simple energy conservation consideration. The dynamics of both
shells is illustrated by pictures of Fig.~2. The splitting of the initial
shell into several separated shells is better seen from pictures of the
radial field energy density $\varepsilon(r)$ shown in Fig.~3. Notice,
that $\varepsilon(r)$ is related with the total energy by
$E=\int{\varepsilon(r)}dr$.

     We also observed analogous splitting of $2\pi$-shell (2) in the model
with $n=3$. In this case the initial $2\pi$-shell splits into two shells
of amplitudes $2\pi/3$ and $4\pi/3$, and some time later $4\pi/3$-shell
splits into a pair of $2\pi/3$-shells. In further evolution all these
$2\pi/3$-shells interact. Corresponding field configurations and radial
energy densities at some typical moments of time are shown in Figs.~4 and 5.

     Notice, that skipped term $\frac{2}{r}\frac{\partial\phi}{\partial r}$
in Eq.~(1) has an essential influence on the evolution of solutions.
In particular, it is responsible for instability of the bubble with respect
to collapse. To study such configurations it is convenient to apply the
effective Lagrangian method. The spherical symmetrical bubble collapse in
the $\lambda\phi^4$-theory was analyzed by this method for the first time
in the thin wall approximation in paper \cite{Zel}, see also \cite{Vol}.
In the nearest future we are planning to discuss the form of effective
Lagrangian and the corresponding equations of motion for multi-shell
configurations studied in the present paper.

     In conclusion it is worth to stress again, that the observed splitting
of the initial shell of DCC bubble of large size and large amplitude leads
to essential prolongation of its mean life-time. That is why the emission
of waves from the region of DCC looks not as an instant but as a rather
long process. The main reason of this prolongation of mean life-time is in
nonlinearity of decay process.

\begin{center}
\bf
Acknowledgments
\end{center}

     This work was supported in part by the Russian Foundation for Basic
Research under grants No~98-02-17316 and No~96-15-96578.
The work of V.~A.~Gani was also supported by the INTAS Grant No~96-0457
within the research program of the International Center for Fundamental
Physics in Moscow.

\newpage

\begin{center}
\bf
Figure captions
\end{center}

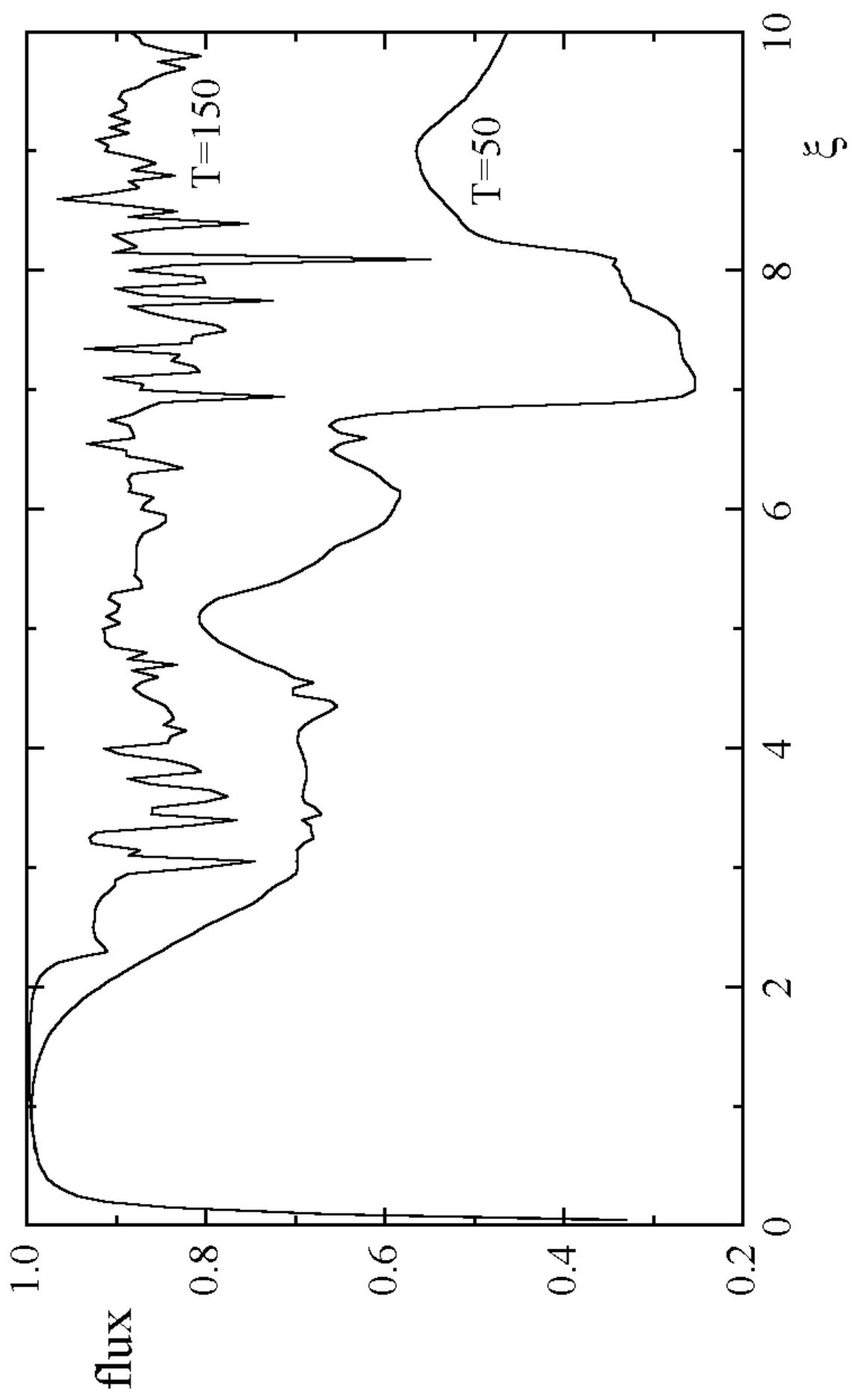
Fig.~1. Energy flux (in units of total energy), that had passed through
the sphere of radius $R=20$ to the moment of time $T$, as a function
of dimensionless parameter $\xi$.

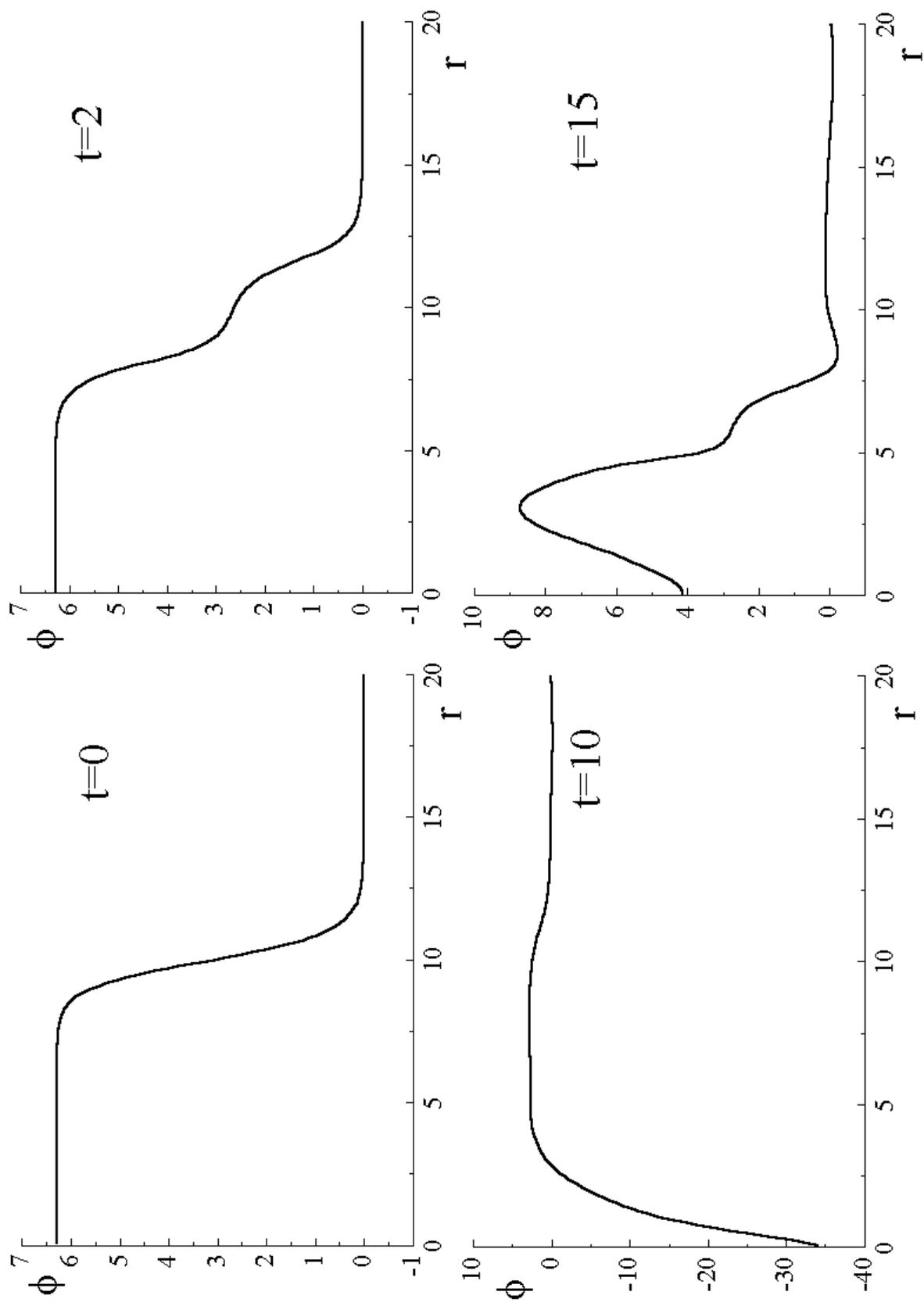
Fig.~2. Typical field configurations at some moments of time $t_i$
in the model with $n=2$.

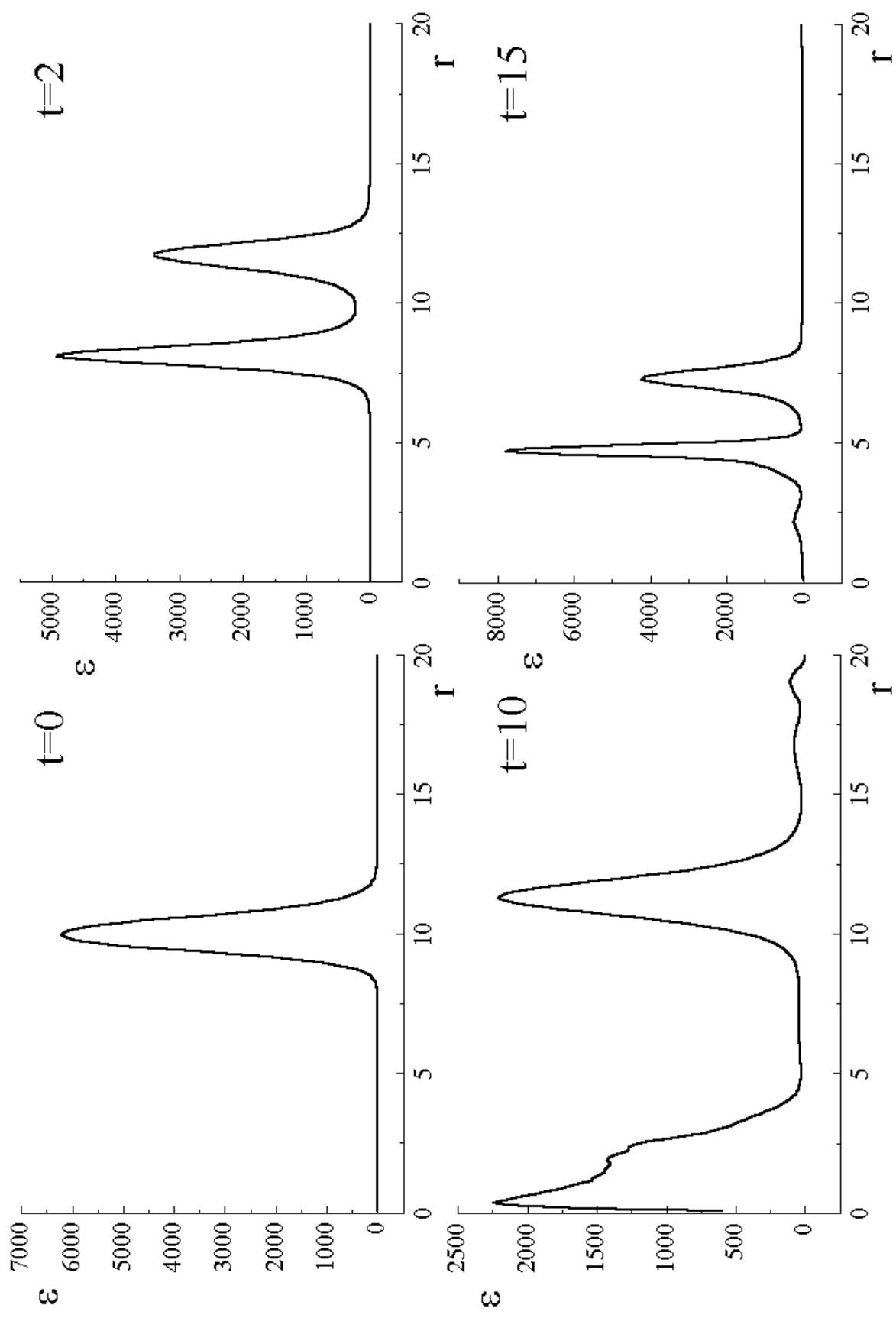
Fig.~3. Radial energy densities corresponding to the field configurations
of Fig.~2.

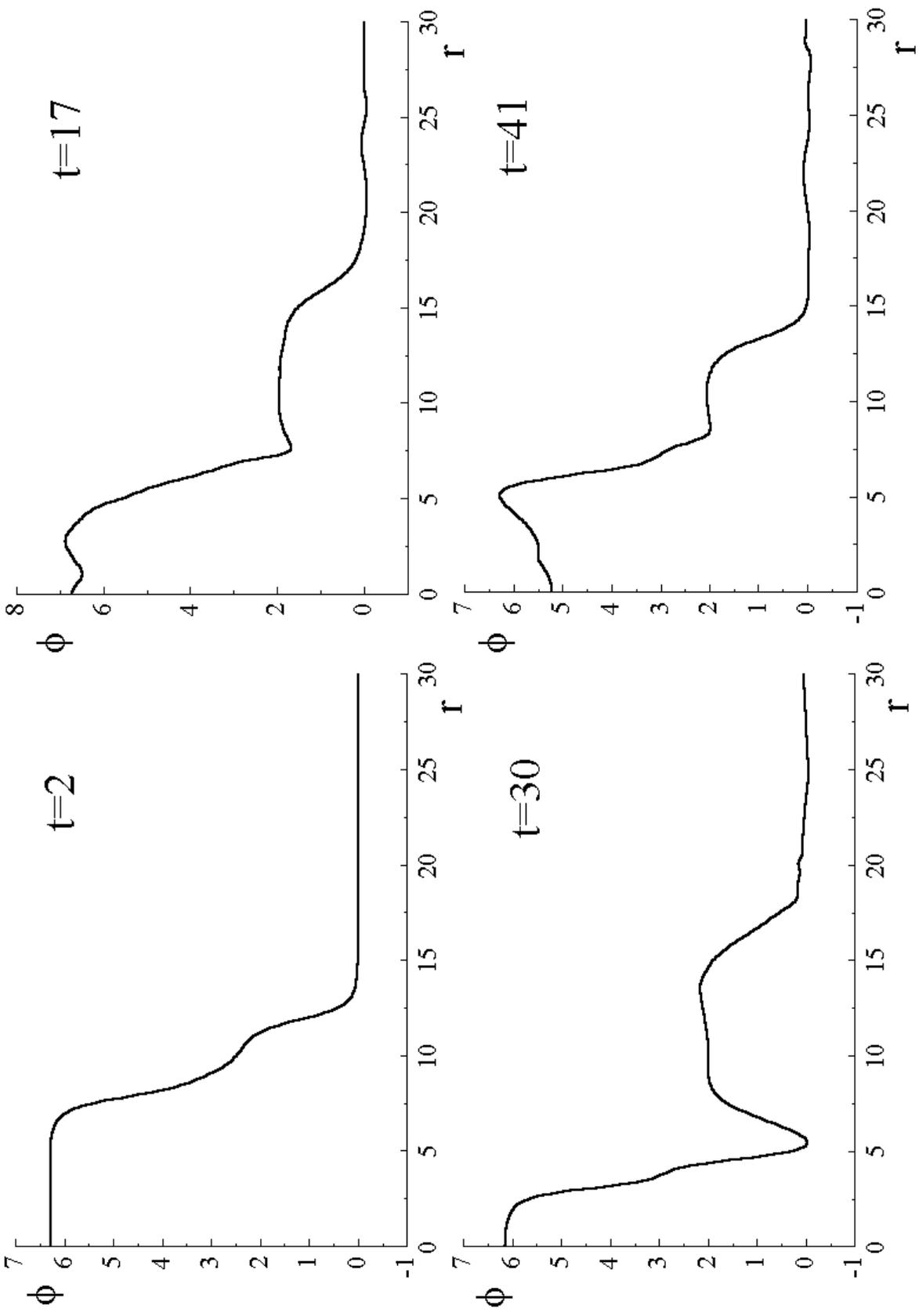
Fig.~4. Typical field configurations at some moments of time $t_i$
in the model with $n=3$.

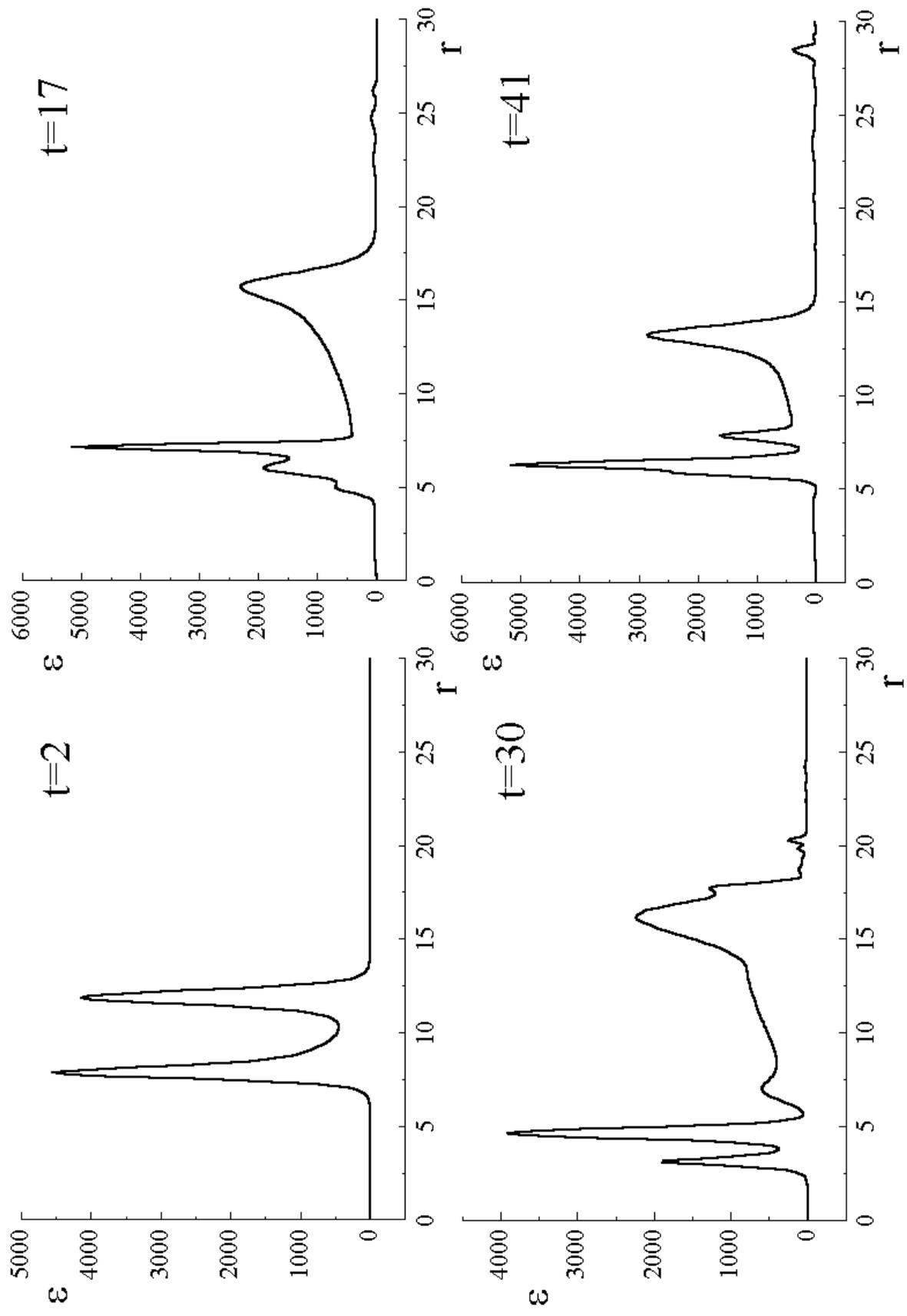
Fig.~5. Radial energy densities corresponding to the field configurations
of Fig.~4.

\end{document}